%% file: Optica-journal-template.tex
\documentclass[9pt,twocolumn,twoside]{opticajnl}
\journal{opticajournal} 

\setboolean{shortarticle}{true}

\usepackage{jabbrv}
\usepackage{lineno}
\usepackage{pdfpages}

\title{Robust inverse design of non-periodic Bloch surface wave sensors}

\author[1]{{Riccardo Bollati}}
\author[2]{Giovanni Pellegrini}
\author[1,**]{Paolo Biagioni}
\author[1,*]{{Jonathan Barolak}}

\affil[1]{Dipartimento di Fisica, Politecnico di Milano, 20133 Milano, Italy}
\affil[2]{Dipartimento di Fisica, Universit\`a degli Studi di Pavia, 27100 Pavia, Italy}

\affil[*]{jjbarolak@gmail.com}
\affil[**]{paolo.biagioni@polimi.it}

\begin{abstract}
One-dimensional photonic crystals supporting Bloch surface waves (BSWs) have emerged as highly sensitive platforms for label-free refractometric sensing. However, the practical performance of these sensors is limited by their vulnerability to fabrication-induced thickness variations. In this Letter, we address this challenge by introducing a robust inverse design framework based on multi-objective genetic optimization. This framework jointly maximizes sensitivity and resilience against manufacturing errors, demonstrating that non-periodic multilayers achieve inherently superior sensitivity–robustness trade-offs, with a pronounced gain in device resilience over conventional periodic designs. By analyzing the optimized designs, we uncover an intrinsic compensation mechanism that physically governs the trade-off and effectively stabilizes the sensor response against structural perturbations. By unlocking the expanded design space of non-periodic multilayers, our optimization strategy paves the way for the practical realization of next-generation BSW sensors.

\end{abstract}

\setboolean{displaycopyright}{false} 

\begin{document}

\maketitle

\section{Introduction}
Bloch surface waves (BSWs) excited at the interface of one-dimensional photonic crystals (1DPCs) have emerged as a powerful platform for refractometric sensing \cite{Michelotti:25}. 1DPCs are dielectric multilayers defined by a step-wise modulation of the refractive index, which breaks continuous translational symmetry along one spatial axis, opening a  photonic bandgap \cite{joannopoulos2008molding}.  The latter, combined with total internal reflection, provides the dual confinement mechanism sustaining BSWs at the 1DPC-superstrate interface. 
\begin{figure}[htbp]
    \centering
    \def\svgwidth{\columnwidth} 
    
    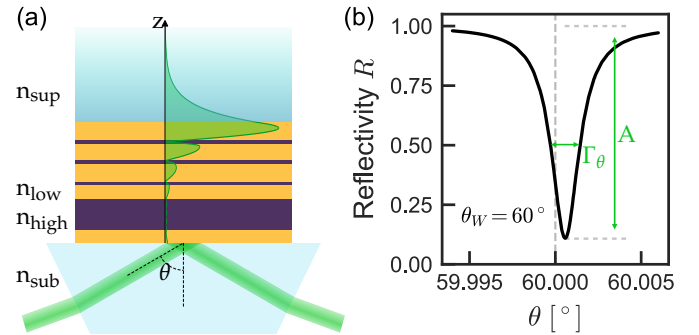
    
    \caption{(a) Schematic of a 1DPC supporting a BSW at the interface with the superstrate, highlighting the characteristic bi-evanescent field profile (green). (b) Angular reflectance profile exhibiting the narrow optical resonance characteristic of BSW sensors, defined by its amplitude ($A$) and linewidth ($\Gamma_\theta$). The working angle $\theta_\text{W}$ is marked by a vertical dashed line.}
    \label{fig:schematic}
\end{figure}
The BSW dispersion relation lies simultaneously below the light line and within the photonic bandgap, resulting in the characteristic evanescent decay of the field away from the interface (Fig. \ref{fig:schematic}a) \cite{joannopoulos2008molding}.
The spatial overlap of this evanescent tail with the superstrate enables highly sensitive, label-free detection of local refractive index variations, allowing for the quantification of target analytes \cite{sinibaldiBlochSurfaceWaves2017,rizzoBlochSurfaceWave2018,barolak2026polarization,aurelioElectromagneticFieldEnhancement2017a} by tracking the shifts of the narrow optical resonance (Fig. \ref{fig:schematic}b).

However, the optical properties of truncated 1DPCs, such as BSW resonances, are intrinsically sensitive to fabrication-induced thickness deviations, since they rely on the precise interference of multiple reflections within the multilayer  \cite{sozuerRobustnessOnedimensionalPhotonic2005,liuDatadrivenInvestigationThickness2024,liewOpticalProperties1D2010}. The impact of fabrication imperfections on BSW resonances and their refractometric sensitivities has so far been assessed only a posteriori on previously optimized structures \cite{Kaliteevski2006,munzert2017multilayer}. Anopchenko \textit{et al.} \cite{anopchenko2016effect} showed that, from a band-structure perspective, modes lying well inside the photonic bandgap exhibit inherently greater robustness. Notably, optimizing for high sensitivity tends to naturally yield modes with this property. Nevertheless, this synergy remains valid only within a limited regime; pushing for ultimate sensitivity requires even stronger field confinement, which conversely amplifies the impact of fabrication-induced perturbations in the layers where the field is localized, as demonstrated by Lereu \textit{et al.} \cite{lereu2019sensitivity}. This intrinsic conflict between performance and fabrication tolerance reflects a broader challenge in nanophotonics \cite{schul2026inverse}. To overcome this fundamental conflict, the field is shifting from passive a posteriori evaluations toward proactive, 'fabrication-aware' strategies. Robust Inverse Design (RID) approaches, already adopted across a wide range of nanophotonic platforms, address this issue by incorporating fabrication variability directly into the optimization process, thereby enabling the identification of structures that maintain their performance under realistic manufacturing conditions \cite{elsawy2021optimization,panda2020robust,schmitt2019optimization}.

While RID has proven highly effective for nanophotonic devices with a large number of geometrical degrees of freedom \cite{elsawy2021optimization}, its effectiveness when applied to conventional periodic 1DPC sensors may be inherently limited by the reduced dimensionality of the design space. Standard 1DPC architectures, defined by a repeating unit cell and a termination layer, present only a few independent design parameters \cite{Michelotti:25,michelotti2013probing}. Recent studies have shown that lifting this periodicity constraint grants non-periodic multilayer configurations access to a significantly richer design space, enabling novel optical functionalities \cite{barolakLeveragingLowIndex2026,shiOptimizationMultilayerOptical2018,jinTransmissiveNonlocalMultilayer2021}. Motivated by these considerations, here we exploit non-periodic multilayers to fundamentally enhance the robustness of 1DPC-based sensors. We adopt a RID approach based on multi-objective genetic optimization, jointly maximizing refractometric sensitivity and resilience against fabrication-induced thickness variations. This evolutionary framework identifies a set of optimal designs representing the best available trade-offs between the conflicting objectives, defining the so-called Pareto front. The high-dimensional optimization is made tractable by the computational efficiency of the 1DPC modeling via the transfer matrix formalism \cite{luce2022tmm}. Evaluating the Pareto fronts resulting from our analysis, we compare periodic and non-periodic architectures, demonstrating that non-periodic designs achieve inherently superior sensitivity–robustness trade-offs. By analyzing the optimized structures, we uncover the physical origin of the sensitivity–robustness trade-off, providing deeper insight into the intrinsic properties of robust BSW sensors. 
These results pave the way for reliable, mass-producible BSW platforms, accelerating their translation from laboratory to real-world applications. Beyond sensing, advanced 1DPC applications such as chiral sensing, spatial filtering, and nonlinear light generation \cite{pellegriniChiralSurfaceWaves2017, liu_single_2022, nkonopskyPhasematchedThirdharmonicGeneration2016} can benefit from our design methodology for achieving fabrication-robust, high-performance devices.

\section{Optimization Framework}

We consider a 1DPC supporting TE-polarized BSWs in the visible regime ($\lambda = 550$ nm), excited by a plane wave incident via a BK7 glass prism ($n_{\text{sub}} = 1.57$) in a Kretschmann configuration, operating in an aqueous environment ($n_{\text{sup}} = 1.33$) at an incidence angle of $60^\circ$, as shown in the schematic in Fig. \ref{fig:schematic}a. The multilayer is comprised of alternating high-index $\text{Ta}_2\text{O}_5$ ($n_\text{H} = 2.15 + i10^{-4}$) and low-index $\text{SiO}_2$ ($n_\text{L} = 1.45 + i10^{-5}$) layers with an $\text{SiO}_2$ termination layer. The thickness of each layer is treated as an independent continuous variable, bounded between 10 and 400 nm to ensure fabrication feasibility. These structural constraints, combined with the simultaneous evaluation of two Figures of Merit (FOMs), addressing refractometric sensitivity and fabrication robustness, define the fitness landscape of our evolutionary search.

The first objective is the maximization of the intensity sensitivity, defined as:
\begin{equation}
S_\text{I} = \left| \frac{d R}{d n_{\text{sup}}} \right| \quad,
\end{equation}
where $R$ is the reflectivity and $n_{\text{sup}}$ is the refractive index of the sensing medium. The reflectivity spectrum exhibits a characteristic Lorentzian dip associated with the BSW resonance, as illustrated in the angular domain in Fig. \ref{fig:schematic}b. $S_\text{I}$ is maximized at the point of steepest slope of this resonance curve, $\theta_\text{W}$ in Fig. \ref{fig:schematic}b,  where small variations in $n_{\text{sup}}$ induce the largest change in reflectivity.  
Whether interrogated in the angular or spectral domain, the sensitivity $S_\text{I}$ at this maximum slope is intrinsically proportional to the standard BSW FOM, employed in the literature (see Supplement 1)\cite{anopchenko2016effect,rizzo2014optimization,ge2022highly}: 

\begin{equation}
    \text{FOM}_{\text{BSW}} = \underbrace{\frac{A}{\Gamma_{\xi}}}_{L_\xi} \underbrace{\frac{d\xi_{\text{res}}}{dn_{\text{sup}}}}_{S_\xi} \propto S_{\text{I}} \quad ,
    \label{eq:decomposition}
\end{equation}
where $\xi$ denotes the generalized interrogation variable (e.g., angle $\theta$ or wavelength $\lambda$). Here, $L_\xi$ is the lineshape factor, determined by the resonance depth $A$ and its linewidth $\Gamma_\xi$, while $S_\xi$ represents the resonance shift induced by variations in the superstrate refractive index. Crucially, adopting  $S_\text{I}$ as optimization FOM offers a substantial computational advantage over $\text{FOM}_\text{BSW}$, as it is evaluated at a single, fixed operating point rather than requiring a full spectral or angular sweep.

The second objective addresses fabrication resilience through a robustness metric, $\text{FOM}_{\text{R}}$, defined as the mean absolute relative error of the intensity sensitivity over a statistical ensemble of thickness-perturbed structures. For each perturbed realization, the resonance condition is allowed to shift within a range of $\pm 1^\circ$ around the nominal working angle, so the sensitivity is re-evaluated at its own operating angle, penalizing the design only for the resulting change in peak sensitivity (see Supplement 2). $\text{FOM}_{\text{R}}$ therefore is computed as:
\begin{equation}
    \text{FOM}_{\text{R}} =
    \frac{1}{N} \sum_{i=1}^{N} \left| \frac{S_{\text{I},i} - S_{\text{I},\text{0}}}{S_{\text{I},\text{0}}} \right|\quad,
    \label{eq:robustness_fom}
\end{equation}
where $S_{\text{I},0}$ is the sensitivity of the nominal structure operating at $\text{60}^\circ$, and $S_{\text{I},i}$ is the sensitivity of the $i$-th perturbed realization at its respective operating angle.
For each candidate 1DPC, an ensemble is generated by applying an independent percentage thickness variation $\delta \sim \mathcal{N}(0, 0.05^2)$ to every layer, modeling a conservative fabrication tolerance of $5\%$ standard deviation. A statistical ensemble of $N=40$ realizations is used per design, balancing reliability of the $\text{FOM}_{\text{R}}$ estimate against computational cost. The minimization of $\text{FOM}_{\text{R}}$ over the ensemble actively drives the genetic algorithm toward designs exhibiting intrinsic resilience to fabrication-induced thickness variations.

\section{Results and Discussion}
To investigate the interplay between sensitivity and robustness, we perform multi-objective optimizations for both periodic and non-periodic 11-layer architectures. The periodic baseline consists of a unit cell repeated 5 times plus a termination layer, yielding only 3 free parameters, while the non-periodic counterpart treats each layer thickness independently. The genetic algorithm was run with a population size of 4000 individuals; further details are provided in Supplement 4. The resulting Pareto fronts, shown in Fig. \ref{fig:paretoFront}(a), reveal a fundamental trade-off between sensitivity and fabrication resilience. Higher sensitivity is inevitably achieved at the cost of reduced robustness. Furthermore, a comparison of the two Pareto fronts reveals that the removal of the periodicity constraint unlocks regions of the performance–robustness landscape entirely inaccessible to periodic 1DPCs, with a $\text{FOM}_{\text{R}}$ improvement above a factor of 2.5 at any matched sensitivity level.

\begin{figure}[H]
    \centering
    \includegraphics[width=1\linewidth]{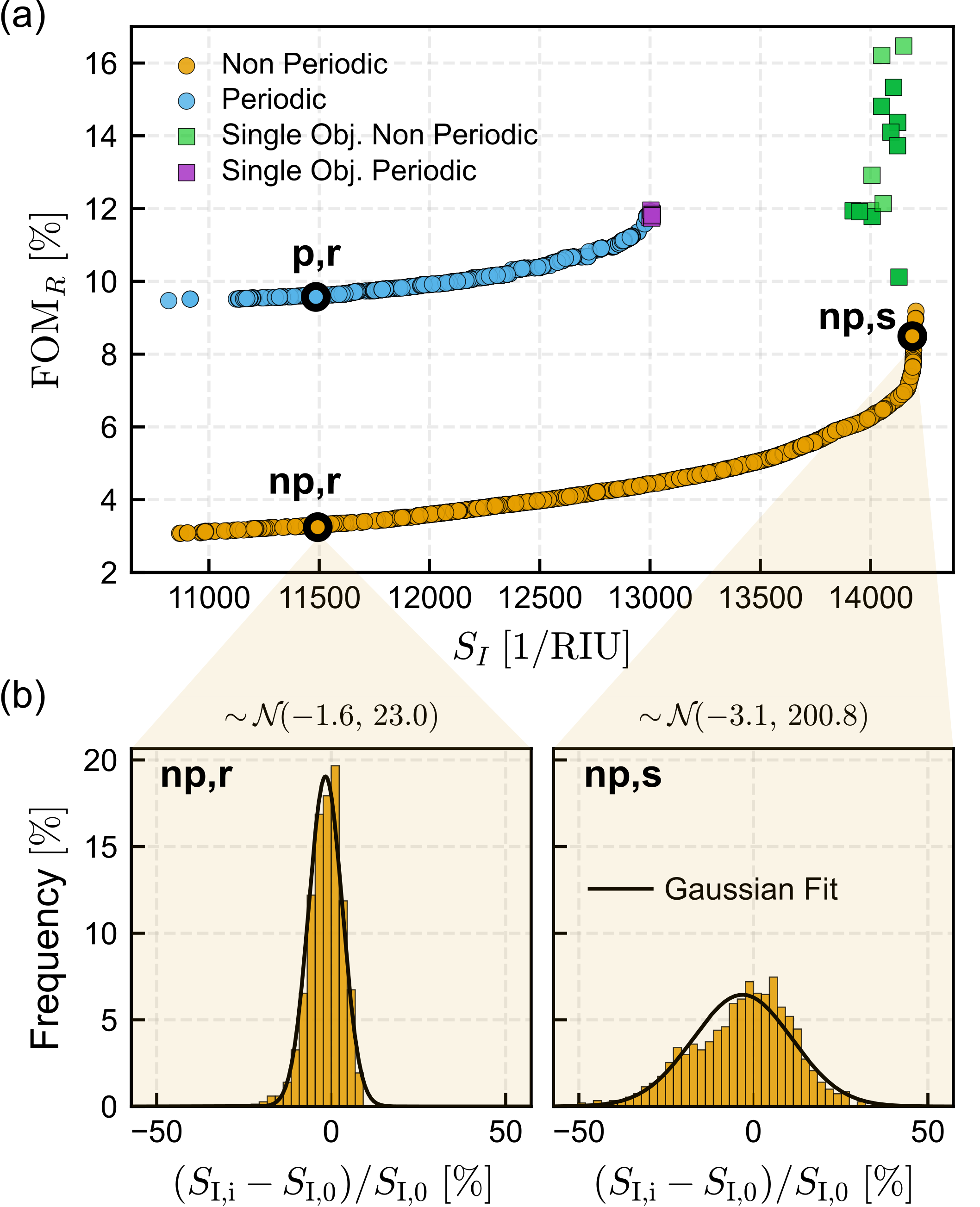}
    
    \caption{(a) Pareto front for 11-layer non-periodic and periodic 1DPCs. Single-objective optimization solutions, targeting exclusively $S_\text{I}$, are indicated as squares. Three representative structures are highlighted: np,r (non-periodic robust, $S_\text{I}~=~11496~\text{RIU}^{-1}$,~$\text{FOM}_\text{R}~=~3.31\%$), np,s (non-periodic sensitive, $S_\text{I}~=~14206~\text{RIU}^{-1}$,~$\text{FOM}_\text{R}~=~9.18\%$), and p,r (periodic robust, $S_\text{I}~=~11500~\text{RIU}^{-1}$,~$\text{FOM}_\text{R}~=~11.25\%$). (b) Statistical distributions of $(S_\text{I,i}~-~S_\text{I,0})/S_\text{I,0}$ over $N~=~1500$ perturbed realizations for np,r and np,s. Gaussian fit parameters are reported above the panels.}
    \label{fig:paretoFront}
\end{figure}

Fig. \ref{fig:paretoFront}(a) also displays single-objective solutions targeting exclusively $S_\text{I}$ maximization (see Supplement 3), converging to the high-sensitivity extreme of each architecture. While the 3-parameter periodic space constrains this extreme to coincide with the Pareto front, the non-periodic single-objective solutions exhibit markedly degraded $\text{FOM}_\text{R}$
at matched sensitivity, underscoring the benefit of the multi-objective approach in the larger non-periodic design space.

To rule out that this trade-off is an artifact of the limited $N = 40$ ensemble used to evaluate $\text{FOM}_\text{R}$, we subject the structures labeled np,r and np,s, drawn respectively from the high-robustness plateau and the high-sensitivity tail (bold-outlined markers in Fig.~2a), to $N = 1500$ independent perturbed realizations. The resulting distributions in Fig. \ref{fig:paretoFront}(b) confirm the contrast, with np,s exhibiting a broad sensitivity dispersion, while np,r yields a sharply confined distribution with a variance reduced by nearly one order of magnitude.
  
\begin{figure}[t]
    \centering
    \includegraphics[width=1\linewidth]{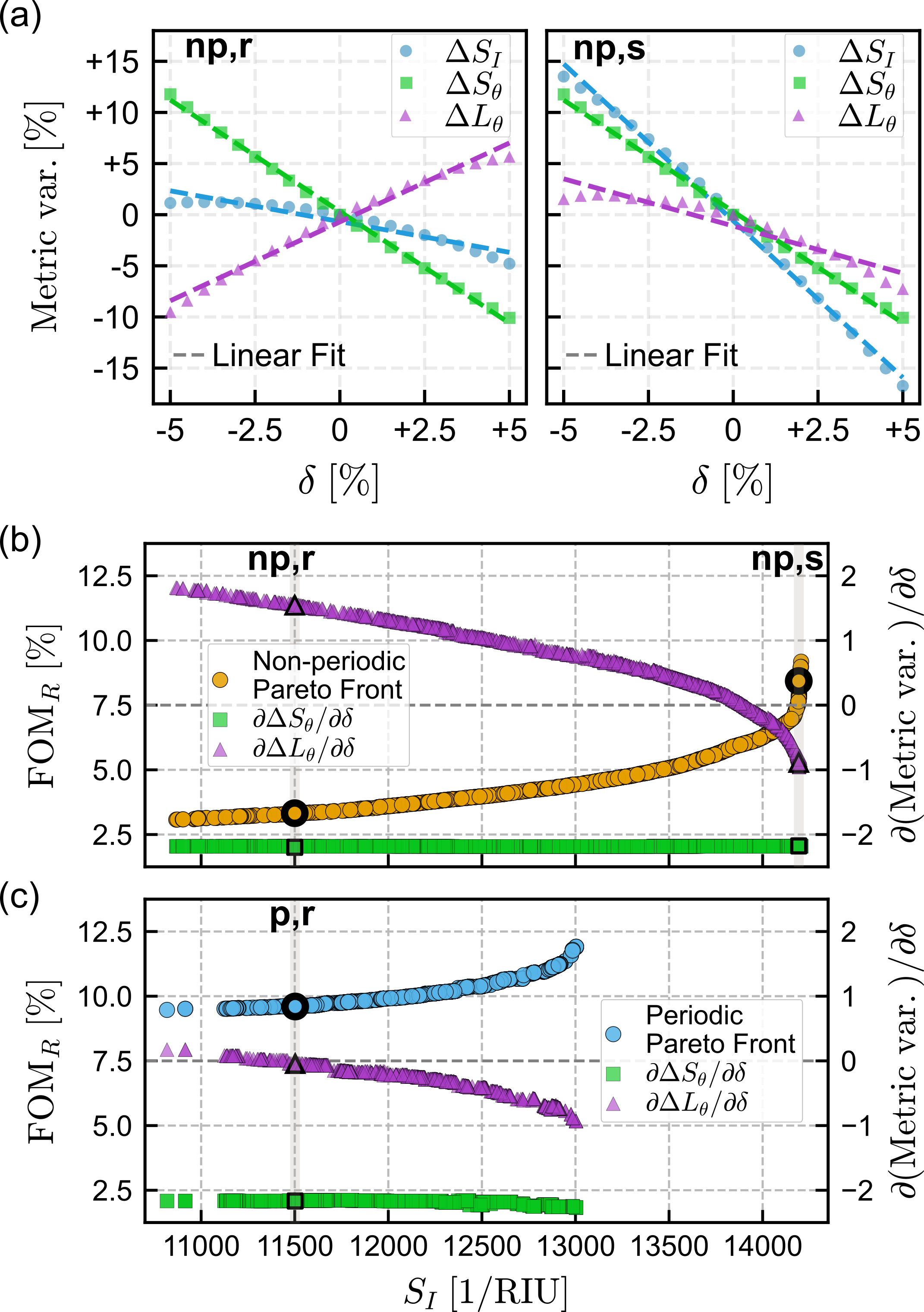}
    
  \caption{(a) Percentage variation of $S_\text{I}$ (blue), 
$L_\theta$ (purple), and $S_\theta$ (green) as a function of top-layer 
thickness perturbation $\delta$ for np,r (left) and np,s 
(right), with linear fits (dashed lines). The fitted slopes are for
np,r: $\partial\Delta S_\text{I}/\partial\delta = -0.60$,\quad 
$\partial\Delta S_\theta/\partial\delta = -2.18$,\quad 
$\partial\Delta L_\theta/\partial\delta = +1.54$;\quad 
for np,s: $\partial\Delta S_\text{I}/\partial\delta = -3.07$,\quad 
$\partial\Delta S_\theta/\partial\delta = -2.18$,\quad 
$\partial\Delta L_\theta/\partial\delta = -0.92$.
(b, c) Perturbation slopes $\partial\Delta L_\theta/\partial\delta$ 
and $\partial\Delta S_\theta/\partial\delta$ along the non-periodic (b) 
and periodic (c) Pareto fronts, as a function of $S_\text{I}$. }
  \label{fig:Slopes}
\end{figure}

To understand the physical origin of this contrasting behavior, we decompose $S_\text{I}$ into the lineshape factor  ($L_\theta$) and the angular sensitivity ($S_\theta$), according to Eq.\ref{eq:decomposition} with $\xi = \theta$, evaluating their variation as the thickness of the topmost layer is varied by $\delta$, with $\delta$ ranging from $-5\%$ to $+5\%$ relative to its nominal value. Being in contact with the superstrate, the topmost layer concentrates the largest fraction of optical power.
The resulting variations of $S_\text{I}$, $S_\theta$ and $L_\theta$ for np,r and np,s are comparatively shown in Fig. \ref{fig:Slopes}(a). 
While the perturbation-induced response of $S_\theta$ is nearly identical between the two designs, the behavior of $L_\theta$ differs fundamentally. In np,s, $L_\theta$  varies in unison with $S_\theta$ , compounding the overall change in $S_\text{I}$. Conversely, in np,r, the opposing trend of $L_\theta$ actively compensates for the variation in  $S_\theta$, keeping the net variation of $S_\text{I}$ remarkably close to zero and effectively stabilizing the sensor response.

Figs. \ref{fig:Slopes}(b–c) extend this analysis along the Pareto fronts, mapping the slopes $\partial S_\theta(\delta)/ \partial\delta$ and $\partial L_\theta(\delta)/ \partial\delta$ for the same topmost-layer perturbation test. In both the non-periodic and periodic fronts, the angular sensitivity slope remains practically constant across all structures. Conversely, the lineshape factor slope undergoes a systematic transition along the non-periodic front (Fig. \ref{fig:Slopes}(b)), shifting from positive, compensating values in the robust regime to negative, compounding values in the high-sensitivity regime.
These trends confirm that this intrinsic compensation physically drives the sensitivity–robustness trade-off, with the perturbation response of the topmost layer emerging as a primary contributor to the device's overall resilience against fabrication imperfections, in agreement with Lereu \textit{et al.} \cite{lereu2019sensitivity}.
In the periodic front (Fig. \ref{fig:Slopes}(c)), the $L_\theta$ slope remains close to zero even in the robust plateau, never reaching values sufficient to compensate the variation of $S_\theta$. Periodic stacks are therefore inherently limited, confirming that the expanded non-periodic design space is necessary to unlock this intrinsic compensation and achieve a superior trade-off.

\begin{figure}[htbp]
    \centering
    \includegraphics[width=1\linewidth]{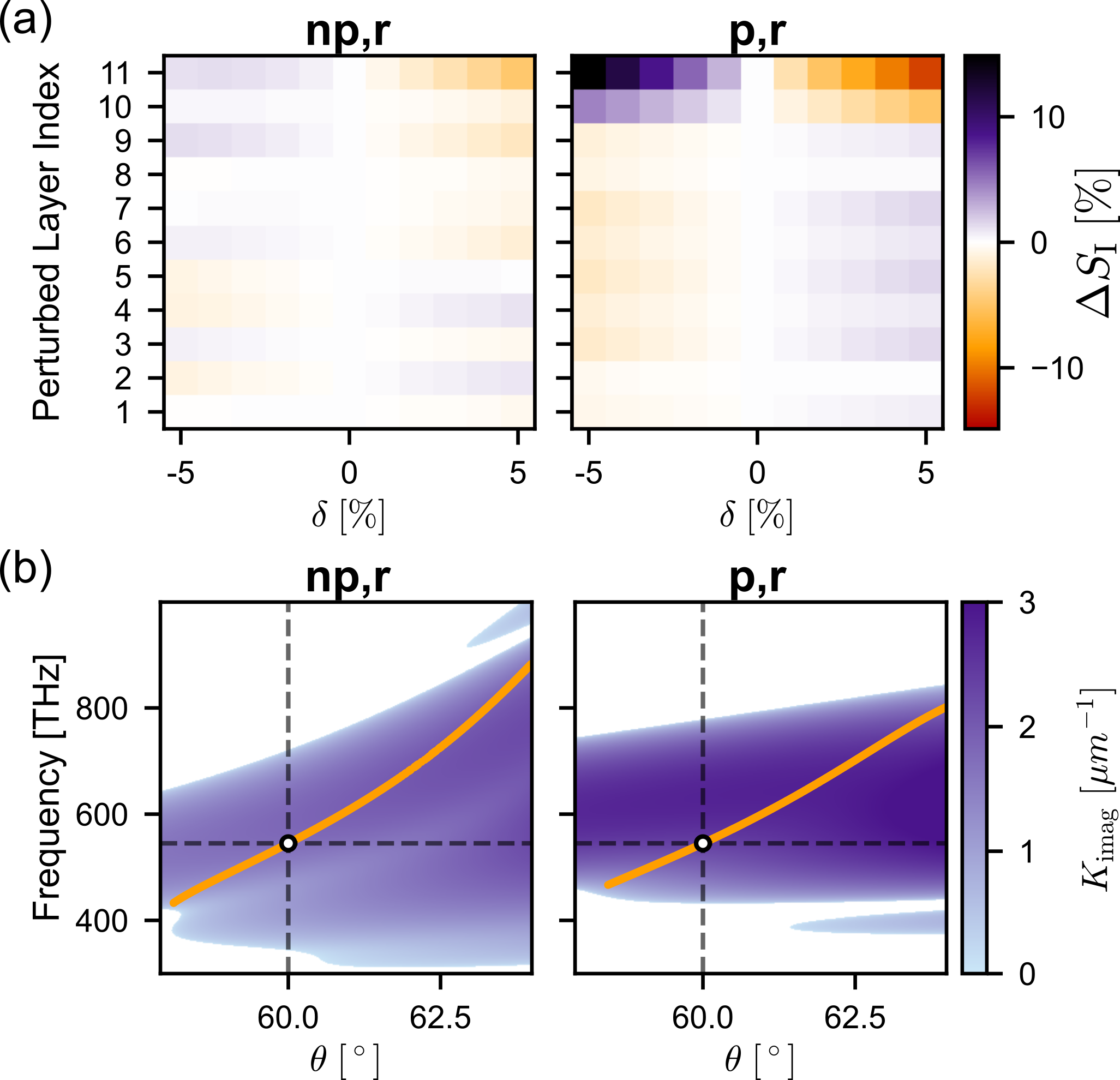}
    
  \caption{(a) Layer-resolved perturbation analysis of the intensity sensitivity ($\Delta S_\text{I}$) as a function of percentage thickness variation $\delta$ for np,r and p,r. (b) Photonic bandgap structure of np,r and p,r, shown as the imaginary part of the Bloch wavevector in the frequency–angle plane; the yellow line represents the BSW dispersion relation. }
  \label{fig:layerPerturbed}
\end{figure}

To further investigate the effect of this compensation mechanism on the layer-resolved vulnerability of the sensor, we generalize the single-layer perturbation test of Fig. \ref{fig:Slopes}(a) to the entire stack, comparing two designs drawn from the robust regimes of the non-periodic and periodic Pareto fronts at nearly identical nominal sensitivities. As shown in Fig.~\ref{fig:layerPerturbed}(a), in the non-periodic design (np,r) the $S_\text{I}$ vulnerability is delocalized almost 
uniformly across the stack, as a consequence of the compensation mechanism occuring in the topmost layer, as discussed above. In contrast p,r exhibits a dominant $S_\text{I}$ vulnerability concentrated in the topmost layers.
Moreover, as shown in Figs. \ref{fig:layerPerturbed}(b), np,r features a significantly broader photonic bandgap than its periodic counterpart (p,r), and its BSW mode also lies further away from the band edges. The band structures are obtained by calculating an effective Bloch wavevector $K_\text{eff}$ for the entire stack, treated as a single macro-unit cell; the forbidden bands, where $K_\text{eff}$ is imaginary and light propagation is evanescent \cite{yeh1977electromagnetic}, define the bandgap. Both observations are consistent with the correlation between bandgap width, mode position, and fabrication robustness previously reported in the literature \cite{Kaliteevski2006,anopchenko2016effect}.

\section{Conclusion}
In this work, we have demonstrated that non-periodic 1DPCs, optimized through a multi-objective genetic framework, achieve superior sensitivity–robustness trade-offs over periodic architectures. The underlying mechanism is an intrinsic compensation: fabrication-induced perturbations in the critical topmost layer drive opposing variations in the lineshape factor and angular sensitivity, which counteract each other to stabilize the overall sensor response, which is proportional to their product. This mechanism is structurally precluded within the constrained parameter space of periodic stacks. These findings directly address the intrinsic conflict between ultimate optical performance and manufacturing tolerances identified as a central open challenge in the field. By shifting from passive, a posteriori, evaluations to a proactive, RID strategy, this work paves the way for the realization of reliable, mass-producible BSW platforms, accelerating the translation of such advanced refractometric sensors from the laboratory to real-world applications.
\bibliography{sample}

\bibliographyfullrefs{sample}



\section*{Supplementary material}

# Robust inverse design of non-periodic Bloch surface wave sensors: Supplemental Document

These supplementary materials provide additional theoretical derivations, a statistical framework for robustness analysis, optimization implementation details, and a convergence analysis of the multi-objective genetic algorithm supporting the results presented in the main text.
Throughout this document, figures and equations from the supplementary material are referred to as Fig. S# and Eq. S#, respectively, while figures and equations from the main text are referred to as Fig. # and Eq. #.
All numerical implementations, together with the structural data underlying the reported plots, are made available in the accompanying open-source GitHub repository. The codes are implemented in Python, relying on *tmm-fast* [1] and *nannos* [2] for the optical simulations, and on *pymoo* [3] for the genetic algorithm optimizations.

## 1. ANALITICAL DERIVATION OF THE RELATIONSHIP BETWEEN $S_I$ AND FOM$_{BSW}$

The experimental interrogation of Bloch surface waves (BSWs) relies on monitoring, in reflection, the variation of the reflectivity spectrum across either wavelength or angle. When the momentum-matching condition is satisfied, the BSW behaves as a resonant mode that temporarily stores electromagnetic energy at the interface before it is dissipated through absorption or radiation leakage [4]. Macroscopically, this phenomenon manifests itself as a pronounced dip in the measured reflectivity spectrum, characterized by a narrow Lorentzian lineshape [5, 6], as shown in Fig. 1b of the main text:

$$R(\xi) = R_{\text{base}} \left( 1 - \frac{A}{1 + \left( \frac{\xi - \xi_{\text{res}}}{\Gamma_\xi / 2} \right)^2} \right) \tag{S1}$$

where $\xi$ represents the interrogation parameter (incidence angle $\theta$ or wavelength $\lambda$), $R_{\text{base}}$ is the off-resonance baseline reflectivity (typically unity due to Total Internal Reflection), $\xi_{\text{res}}$ is the resonance centroid, $A$ is the coupling efficiency ($0 \leq A \leq 1$), and $\Gamma_\xi$ is the Full Width at Half Maximum (FWHM).

In sensing applications, this resonance is typically monitored via angular ($\xi = \theta$) or wavelength ($\xi = \lambda$) interrogation. The fundamental metric describing the sensor's response is the sensitivity $S_\xi = \frac{d\xi_{\text{res}}}{dn_{\text{sup}}}$, which tracks the physical shift of the resonance minimum in the chosen interrogation scheme. To comprehensively assess the resolving power, $S_\xi$ is standardly multiplied by the ratio of the modulation depth to the linewidth ($A/\Gamma_\xi$), referred to as the lineshape factor ($L_\xi$), yielding the conventional Figure of Merit (FOM$_{\text{BSW}}$), see Eq.2 of the main text [7–9].

To establish the analytical relationship connecting these metrics with our intensity sensitivity $S_I$, reported in Eq. 1 of the main text, we assume that the system operates on the linear flank of the BSW reflectivity dip, away from the resonance minimum, where the slope of the curve, and hence the sensitivity, is maximized. Under this condition, and applying the rigid shift approximation, valid for sufficiently small refractive index variations, the entire resonance dip translates rigidly along $\xi$ with $A$ and $\Gamma_\xi$ remaining invariant, so that the index variation $\Delta n_{\text{sup}}$ is entirely encoded as a shift of the resonance centroid $\xi_{\text{res}}$. Under these assumptions, we can write:

$$S_\xi = \frac{d\xi_{\text{res}}}{dn_{\text{sup}}} \approx \frac{d\xi}{dn_{\text{sup}}} \quad . \tag{S2}$$

By applying the chain rule, the variation of reflectivity with respect to the refractive index ($S_I$) can be expressed as the product of $\partial R/\partial \xi$, the derivative of the Lorentzian optical signature of the resonant coupling with respect to $\xi$, and the resonance sensitivity $S_\xi$ introduced in Eq. 2:

$$S_I = \frac{\partial R}{\partial \xi} \cdot \frac{\partial \xi}{\partial n_{\text{sup}}} \quad . \tag{S3}$$

Starting from the Lorentzian definition and differentiating with respect to $\xi$, we obtain:

$$\frac{\partial R}{\partial \xi} = A \cdot \frac{2(\xi - \xi_{\text{res}}) \cdot \left(\frac{\Gamma_\xi}{2}\right)^2}{\left[(\xi - \xi_{\text{res}})^2 + \left(\frac{\Gamma_\xi}{2}\right)^2\right]^2} \quad , \tag{S4}$$

where we have assumed, without loss of generality, that the baseline reflectivity is unity ($R_{\text{base}} = 1$). We set the working point at the maximum slope (the inflection point of the resonance), located at $\xi_{\text{work}} = \xi_{\text{res}} \pm \frac{\Gamma_\xi}{2\sqrt{3}}$. By evaluating the derivative (Eq. S4) at $\xi_{\text{work}}$ and substituting both this result and $S_\xi \approx \partial\xi/\partial n_{\text{sup}}$ (Eq. S2) into the chain rule expression (Eq. S3), we derive the final relationship:

$$S_{\text{I}} = A\frac{3\sqrt{3}}{4\Gamma_\xi} \cdot S_\xi \qquad \rightarrow \qquad S_{\text{I}} \propto \frac{A}{\Gamma_\xi} S_\xi = L_\xi \cdot S_\xi \quad . \tag{S5}$$

This derivation provides the theoretical justification for adopting $S_{\text{I}}$ as a valid surrogate for $\text{FOM}_{\text{BSW}}$. Within the rigid-shift approximation, and irrespective of whether the resonance is interrogated in the angular or spectral domain, the sensitivity $S_{\text{I}}$ evaluated at $\xi_{\text{work}}$ is proportional to the standard BSW Figure of Merit [7–9]:

$$\text{FOM}_{\text{BSW}} = \underbrace{\frac{A}{\Gamma\xi}}_{L\xi} \underbrace{\frac{d\xi_{\text{res}}}{dn_{\text{sup}}}}_{S\xi} \propto S_{\text{I}} \quad . \tag{S6}$$

Here, $L_\xi$ is the lineshape factor, determined by the resonance depth $A$ and its linewidth $\Gamma_\xi$ (see Eq.2). Crucially, adopting $S_{\text{I}}$ rather than $\text{FOM}_{\text{BSW}}$ as the optimization objective offers a substantial computational advantage, as it is evaluated at a single, fixed operating point ($\xi_{\text{work}}$) rather than requiring a full spectral or angular sweep to extract $A$, $\Gamma_\xi$, and the computation of multiple reflectivity spectra at different $n_{\text{sup}}$ to track the resonance shift and obtain $S_\xi$.

### A. Numerical evaluation of $S_{\text{I}}$

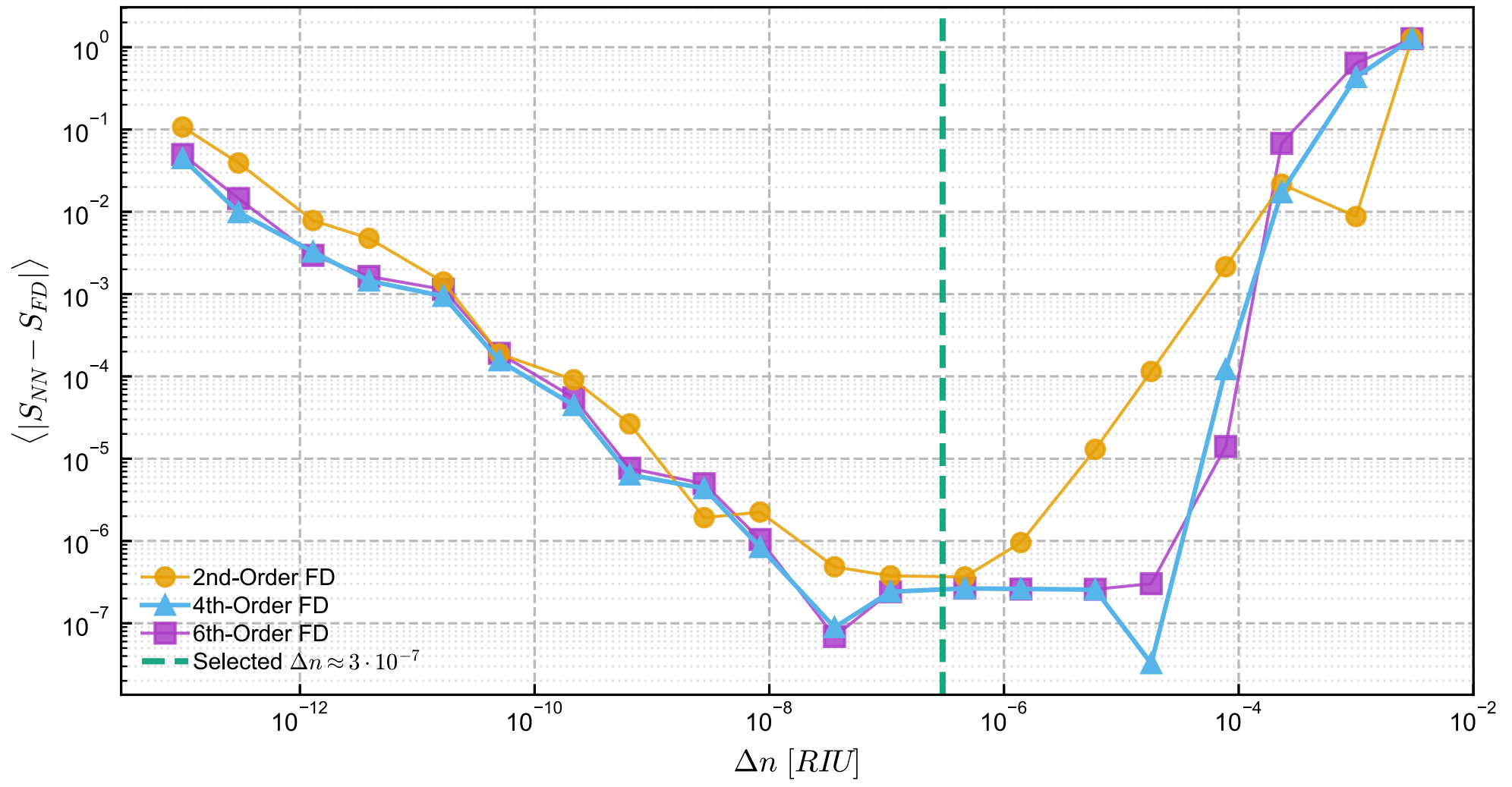


**Fig. S1.** Mean absolute error between the finite-difference sensitivity $S_{\text{FD}}$ and the reference sensitivity $S_{\text{NN}}$ as a function of the refractive-index perturbation $\Delta n$, for second-, fourth-, and sixth-order finite-difference schemes. The selected perturbation step, $\delta n \simeq 3 \cdot 10^{-7}$, indicated by the dashed vertical line.

Since *tmm_fast* [1] does not support automatic differentiation with respect to the refractive index, $S_{\text{I}}$ is computed via a fourth-order central finite-difference scheme applied to the reflectivity at

the superstrate, evaluated at $\xi_{\text{work}}$. To preserve the computational efficiency of *tmm_fast*, the perturbed reflectivity evaluations required by the finite-difference stencil are computed simultaneously through a vectorized batching strategy, avoiding explicit loops over the perturbation values.

The order of the finite-difference scheme and the perturbation step $\delta n$ were selected by comparing the resulting sensitivity against the analytical derivatives provided by *nannos* [2], used here as a reference benchmark, as its analytical formulation comes at a significantly higher computational cost than *tmm_fast*. This validation was carried out on a high-sensitivity, single-objective optimized structure (see section S3), characterized by a particularly sharp resonance, in order to assess the robustness of the finite-difference approach under the most demanding spectral conditions. The mean absolute error between the finite-difference and reference sensitivities was evaluated as a function of $\delta n$ for second-, fourth-, and sixth-order schemes (Fig. S1). This error was computed over a narrow angular range tightly localized around the resonance dip. All schemes display a characteristic U-shaped error curve, reflecting the trade-off between truncation error at large $\delta n$ and numerical noise at small $\delta n$, while higher-order schemes additionally exhibit a plateau region where the error becomes weakly dependent on $\delta n$. The fourth-order scheme was selected as it shows a wider and more reliable plateau than the second-order scheme, whereas the sixth-order scheme provides no significant improvement within this region despite a larger stencil and higher computational cost. The perturbation step was accordingly fixed at $\delta n \simeq 3 \times 10^{-7}$, lying within the plateau of the fourth-order curve and ensuring a stable and reliable sensitivity estimate even for sharp resonant features.

## 2. STATISTICAL FRAMEWORK FOR ROBUSTNESS

Layer thicknesses inevitably deviate from their nominal values during fabrication. Whether these deviations critically degrade sensor performance, however, depends on how the structure responds to them, through two distinct and concurrent mechanisms.

In the following analysis, consistent with the $\text{FOM}_\text{R}$ evaluation reported in the main text, the wavelength is kept fixed and the statistical robustness framework is developed in terms of the angular working point $\theta_\text{W}$; an analogous treatment could equivalently be carried out in the spectral domain, fixing the angle and analyzing the working wavelength.

First, thickness deviations from nominal values generally perturb the interference conditions within the multilayer stack, shifting the BSW resonance to a new optimal working angle $\theta_\text{W}$ [7, 10]. This shift originates from the perturbation-induced change in the round-trip phase accumulated by the field within the stack: its magnitude is therefore expected to scale with both the magnitude of the thickness deviation and the fraction of the field locally confined within the perturbed layer, while its direction is set by the sign of the thickness deviation. Since, across all structures considered in this work, the topmost layer concentrates the largest fraction of optical power, it is expected to dominate this effect. Furthermore, $S_\text{I}$ depends strongly on the operating angle, particularly in the vicinity of the total internal reflection condition; consequently, this shift alone directly impacts the achievable sensitivity. Second, the modified stack would generally not constitute the optimal design for that operating condition, and its sensitivity therefore falls somewhat short of the maximum achievable for that $\theta_\text{W}$.

To illustrate the combined effect of these two mechanisms, Fig. S2a reports the peak sensitivity $S_{\text{I},i}$ against the corresponding optimal working angle $\theta_\text{W}$ for $N = 1500$ perturbed realizations of a 1DPC optimized through single-objective optimization, targeting $S_\text{I}$ exclusively as detailed in Section 3. The colormap encodes the percentage variation of the topmost layer $\delta$. As reported in the main text, each perturbed realization is generated by independently perturbing every layer thickness by a relative deviation $\delta$ drawn from a zero-mean Gaussian distribution, $\delta \sim \mathcal{N}(0, 0.05^2)$, modeling a conservative fabrication tolerance of 5% standard deviation. The single-objective design (SOD) considered in this analysis corresponds to the green square in Fig. 2a, located at $S_{\text{I},0} = 14058.9$ RIU$^{-1}$ and $\text{FOM}_R = 12.14\%$. Fig. S2b recasts the same ensemble as a one-dimensional distribution, collapsing the angular dependence and expressing the vertical axis as $(S_{\text{I},i} - S_{\text{I},0})/S_{\text{I},0}$, thereby allowing a more direct, quantitative comparison with Fig. 2b of the main text.

As shown in Fig. S2a, the majority of perturbed realizations lie along a band that reflects the steep dependence of $S_\text{I}$ on $\theta_\text{W}$, consistent with the first mechanism discussed above. This band, however, is not a sharp line but rather a region of finite width, with points sharing a similar $\theta_\text{W}$ nonetheless spanning a range of $S_{\text{I},i}$ values: this width is precisely the signature of the second mechanism, namely the sub-optimality of each perturbed configuration relative to the

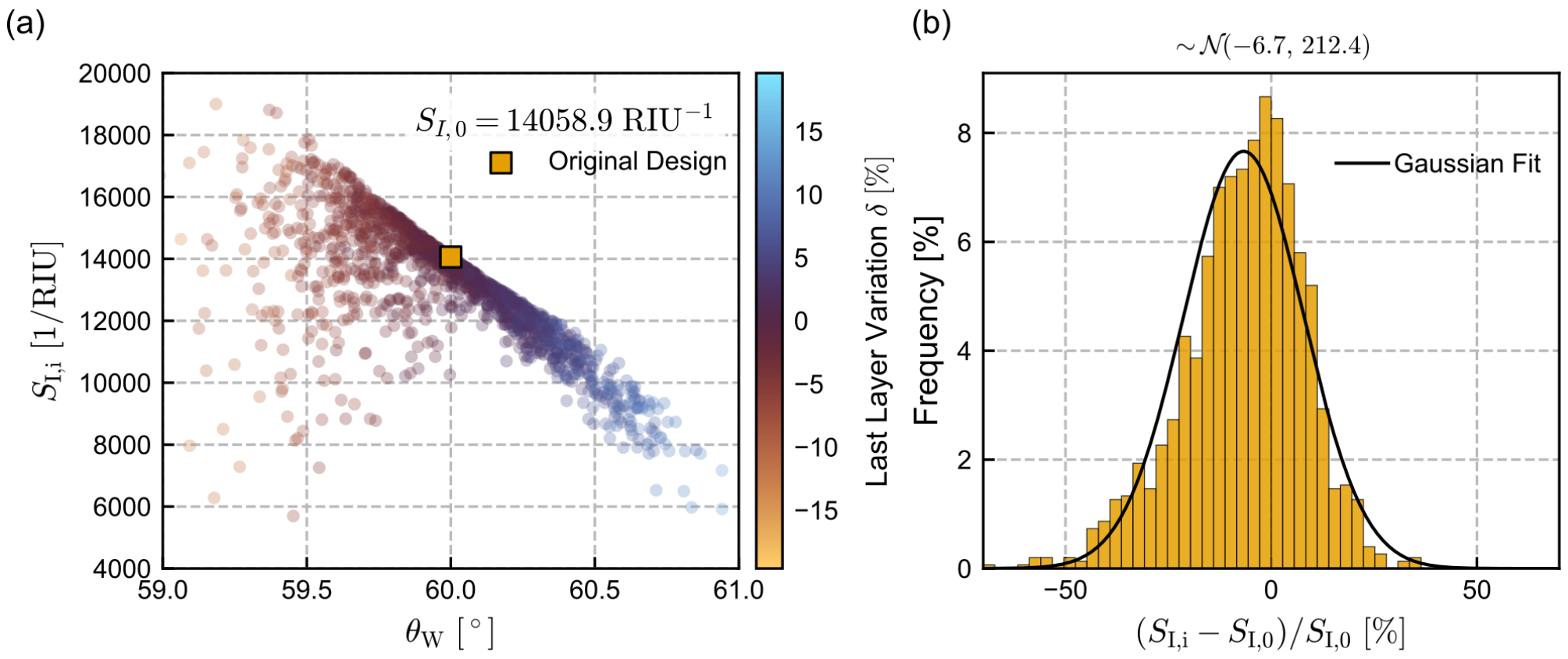


**Fig. S2.** Statistical analysis of the SOD (green square in Fig. 2a),over $N = 1500$ perturbed realizations. (a) Peak sensitivity $S_{I,i}$ versus the corresponding optimal working angle $\theta_W$, with the colormap encoding the relative thickness deviation $\delta$ of the topmost layer; the nominal unperturbed design is indicated by the square marker. (b) Statistical distribution of $(S_{I,i} - S_{I,0})/S_{I,0}$, with the solid line representing a Gaussian fit; Gaussian fit parameters are reported above the panel.

best design achievable at the specific working angle. While most realizations remain relatively close to this band, a non-negligible fraction, predominantly at $\theta_W < 60°$, departs markedly below it, undergoing a substantially more severe degradation in sensitivity. This combined behavior translates, in Fig. S2b, into a distribution whose mean is shifted from zero by $-6.7\%$, a considerable systematic offset indicating that the perturbed ensemble is, on average, biased toward lower sensitivity.

Building on this evidence, an effective robustness metric should drive the optimizer toward structures whose sensitivity, under fabrication-induced perturbations, deviates as little as possible from its nominal value. We deliberately disregard the shift in the working angle itself, since re-aligning the interrogation scheme to a new $\theta_W$ is, experimentally, a minor adjustment. The metric should instead exclusively target the resulting variation in peak sensitivity, evaluated at each realization's own optimal working angle.

A first option, common in robust inverse design, is worst-case (min-max) optimization [11], where only the most degraded realization is penalized. This approach is not suited to our problem for two reasons. First, the worst-case realization cannot be identified a priori.Second, penalizing only the single worst realization ignores the statistical distribution of the ensemble, providing no information on how sensitivity changes across the full set of perturbed configurations, failing to jointly capture the effects of both the resonance shift and the sub-optimality mechanisms discussed above. A second option is to treat robustness statistically, using the variance of the perturbed sensitivities as the optimization objective, as in Elsawy *et al.* [12]. Variance reduces dispersion, but it measures this dispersion around the mean of the perturbed ensemble, not around the nominal sensitivity $S_{I,0}$. A design could therefore become very stable while drifting away from the performance it was originally optimized for.

To prevent this drift, an alternative is to evaluate the error relative to the nominal target using the Mean Squared Error, as in Panda *et al.* [13], or equivalently its square root. While these metrics correctly anchor the penalty to $S_{I,0}$, their quadratic formulation is not well suited for our limited ensemble of $N = 40$ perturbed structures: a small number of severely degraded realizations can dominate the metric, causing the optimizer to reject designs that are statistically robust across the majority of the ensemble. We instead want every realization to contribute proportionally to the total cost, regardless of the magnitude of its deviation.

This is why we instead define $\mathrm{FOM_R}$ as the relative Mean Absolute Error with respect to $S_{I,0}$ (Eq. 3 of the main text, reported here for convenience):

$$\mathrm{FOM_R} = \frac{1}{N}\sum_{i=1}^{N}\left|\frac{S_{I,i} - S_{I,0}}{S_{I,0}}\right| \quad , \tag{S7}$$

Being linear and symmetric, $\mathrm{FOM_R}$ weighs every realization equally and penalizes deviations

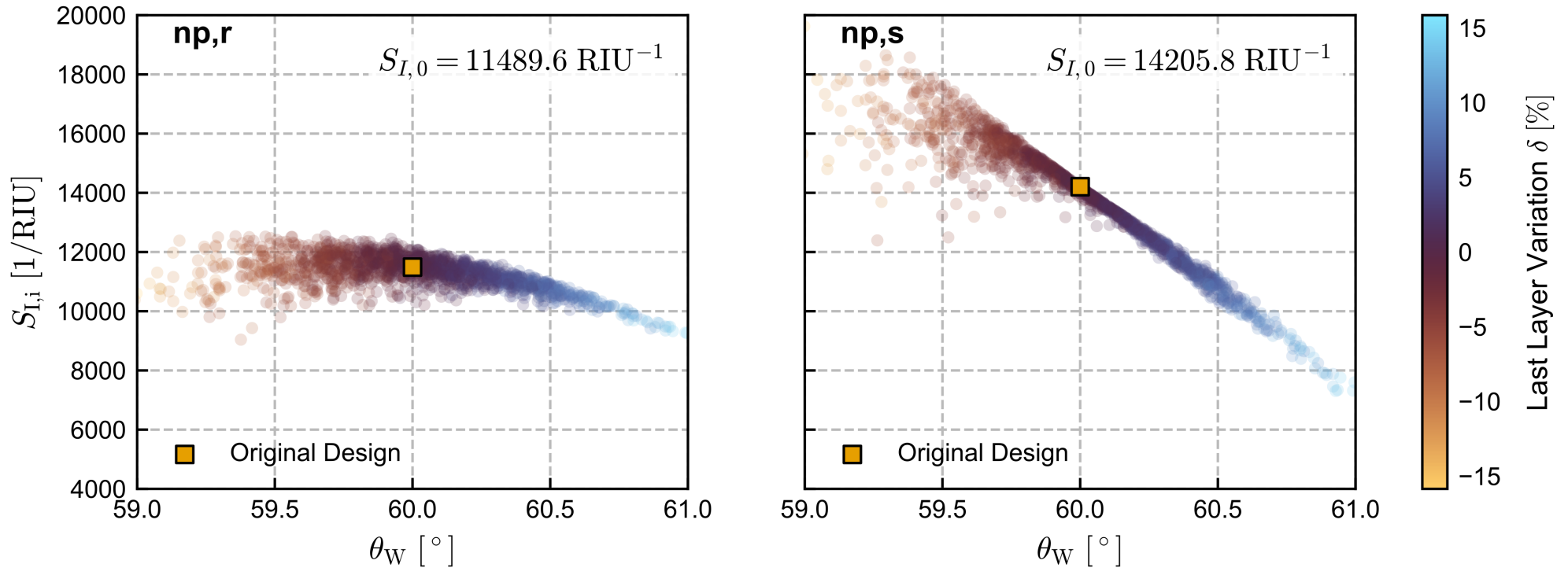


**Fig. S3.** Statistical analysis of the robust design (np,r, left) and the high-sensitivity design (np,s, right) over $N = 1500$ perturbed realizations. Peak sensitivity $S_{\mathrm{I},i}$ versus the corresponding optimal working angle $\theta_\mathrm{W}$, with the colormap encoding the relative thickness deviation $\delta$ of the topmost layer; the nominal unperturbed design is indicated by the square marker.

in either direction with respect to $S_{\mathrm{I},0}$. As a direct consequence, minimizing it simultaneously reduces the spread of the perturbed sensitivities and keeps their average close to the nominal value, addressing both issues discussed above. It also has a simple physical reading: it is the average percentage variation of the sensitivity under fabrication-induced perturbations.

Building on the analysis of Fig. 2b of the main text, Fig. S3 extends it by plotting the peak sensitivity $S_{\mathrm{I},i}$ as a function of the corresponding optimal working angle $\theta_\mathrm{W}$, with the colormap encoding the percentage thickness variation of the topmost layer, following the same representation already adopted in Fig. S2a. This allows a direct comparison between np,r, np,s and the SOD discussed above.

A first observation concerns the mean of the Gaussian fit to the perturbed distributions of the high-sensitivity designs, np,s and the SOD. Despite their comparable nominal sensitivity, the mean offset is reduced for np,s (−3.1%) compared to the SOD (−6.7%), already pointing to a beneficial effect of the multi-objective optimization even at the high-sensitivity end of the Pareto front. Comparing the three designs in Figs. S2a and S3, the slope of the $S_{\mathrm{I},i}$–$\theta_\mathrm{W}$ trend is essentially the same for the SOD and np,s, indicating that the multi-objective optimization does not act on the angular dependence itself at the high sensitivity end of the Pareto front. What changes is instead the width of the band around this trend: relative to the SOD, this band is considerably narrower for np,s, and the worst-case realizations, those departing most severely from the main trend, are markedly less frequent and less extreme. Moving toward the robust end of the Pareto front, this slope is instead markedly reduced for np,r, signaling a fundamentally different angular dependence.

In np,s and the SOD, the steep dependence of $S_\mathrm{I}$ on the resonance angle dominates the total variation, with even modest angular displacements translating into large sensitivity penalties. In np,r, by contrast, this angular dependence is substantially flattened: for $\theta_\mathrm{W} > 60°$ the trend remains close to linear but with a much gentler slope, while for $\theta_\mathrm{W} < 60°$ it loses any clear linear character altogether, bending downward toward lower $S_{\mathrm{I},i}$ values rather than continuing to rise. As a result, the overall variance reflects both degradation mechanisms more evenly, rather than being dominated by the angular shift alone. This contrast also appears in the position of the nominal design within its own perturbed ensemble: at the high-sensitivity end, it sits at the very top of the distribution at $\theta_\mathrm{W} = 60°$, with essentially no perturbed realization outperforming it, whereas for np,r it lies at the center of the scattered cloud, in line with the smaller mean offset of the Gaussian fit in Fig. 2b of the main text. This flattening is precisely the signature of $\mathrm{FOM_R}$: by evaluating each perturbed realization at its own optimal working angle and penalizing only the resulting variation in peak sensitivity, the metric directly targets both degradation mechanisms at once. Minimizing $\mathrm{FOM_R}$ as an optimization objective therefore drives the algorithm toward architectures exhibiting this flattened angular dependence, as directly evidenced by the markedly narrower and more stable sensitivity distribution of np,r.

Finally, it is worth noting that, across all three designs, the angular span of the perturbed ensemble and the correlation between the working angle shift and the percentage variation

$\delta$ of the topmost layer remain consistent: positive deviations push $\theta_W$ to larger angles while negative deviations shift it toward smaller values, with the amplitude of the shift scaling with the magnitude of $\delta$.

## 3. SINGLE OBJECTIVE OPTIMIZATION

To address the optimization of the sensitivity of the 1DPC, we adopted Differential Evolution (DE) as the core optimization algorithm. DE is specifically designed for continuous parameter spaces and operates directly on real-valued vectors, making it particularly well suited for photonic design problems where the optimization variables, layer thicknesses in our case, are continuous quantities and the objective landscape is highly non-convex [14]. Beyond its global search capabilities, a distinguishing feature of DE is its algorithmic simplicity: the method relies on a compact set of control parameters, namely the population size, the scaling factor $F$, the crossover rate $CR$, and the mutation strategy. The mutation strategy specifies the rule governing how trial individuals for the next generation are generated from the current population, and, together with the other parameters, eases the tuning process compared to more complex evolutionary frameworks.

However, like all genetic algorithms, DE offers no formal guarantee of having reached the global optimum in non-convex optimization landscapes. Rather than performing a purely random search, DE explores the parameter space through an evolutionary process that combines stochastic operators — mutation and crossover — with a selection mechanism progressively biasing the population toward higher-fitness regions. This makes DE a robust and reliable heuristic approach for global search, yet, since it does not rely on gradient-based stationarity conditions, convergence to the exact global optimum cannot be formally certified, and the final solution may represent only a near-optimal configuration. Furthermore, the quality of the obtained solution can depend on the initial random seed, making the definition of a robust and reproducible optimization routine a critical prerequisite.

Following standard heuristics [14], we set the scaling factor $F = 0.5$ and, given the non-separable nature of the problem, where layer thicknesses are highly interdependent in determining the optical response a high crossover rate $CR = 0.9$ was selected to preserve the structural information carried by the trial individuals [15]. Among the available mutation strategies, we adopted *DE/target-to-best/1/bin* [16] (also referred to in the literature as *rand-to-best/1/bin* [17]). In this strategy, each trial individual is generated by perturbing the current individual with two complementary terms: a directional component that pulls it toward the best solution found so far, and a stochastic difference component, scaled by $F$, computed from two randomly selected individuals in the population. Combining these two terms allows the strategy to balance exploitation, driven by the directional pull toward the best individual, with exploration, introduced by the randomly sampled difference vector. In our preliminary tests, this strategy consistently showed strong convergence toward high-performing solutions, together with a reduced dependence on the random seed compared to alternative mutation variants, motivating its adoption throughout this work. The selected optimization routine employs the *DE/target-to-best/1/bin* strategy with a population size of 9000 individuals evolved over 4000 generations, substantially larger than the standard heuristic of $10 \cdot D$ [14], in order to suppress stochastic fluctuations and minimize sensitivity to random initialization, mitigating the risk of genetic drift [14]. This configuration consistently converges toward the same solution across independent optimizations. Defining the coefficient of variation as $CV = \sigma / \mu$, where $\mu$ and $\sigma$ denote the mean and standard deviation of the optimized $S_I$ across independent runs initialized with different random seeds. The resulting $CV$ does not exceed 2%, confirming the robustness of the adopted routine.

## 4. MULTI-OBJECTIVE OPTIMIZATION

To address the multi-objective optimization of $S_I$ and $FOM_R$, we employed the Non-dominated Sorting Genetic Algorithm II (NSGA-II), developed by Deb *et al.* [18] and implemented via the *pymoo* library [3]. NSGA-II was selected as a well-established multi-objective evolutionary algorithm [18, 19] that simultaneously optimizes all objectives without requiring any a priori specification of objective weights or assumptions on the front geometry, making it particularly suited to the present problem where the sensitivity–robustness trade-off structure is not known in advance. From a computational standpoint, NSGA-II achieves a reduced complexity of $\mathcal{O}(MN^2)$, where $M$ is the number of objectives (here $M = 2$) and $N$ is the population size [18]. We adopted a population size of $N = 4000$ individuals to ensure extensive exploration of the high-dimensional

non-periodic design space. Furthermore, NSGA-II relies on a robust elitist mechanism whereby non-dominated solutions are retained across generations: this guarantees that optimal candidates are never discarded due to the stochastic nature of the genetic operators, directly accelerating convergence toward the optimal sensitivity–robustness trade-off. Preserving this elitist property requires a strictly deterministic evaluation of the objective functions. Accordingly, the same fixed set of $N \times L$ percentage thickness variations, where $N = 40$ is the ensemble size and $L = 11$ is the number of layers, was applied to every candidate structure throughout the optimization, guaranteeing a fully deterministic evaluation of $\mathrm{FOM_R}$. While this choice is essential for algorithmic stability, it introduces a potential bias toward the fixed perturbation ensemble; the extent and impact of this bias are addressed by a validation test performed on the final Pareto front (see Figs. S4b, S5b, S6b).

In the adopted algorithm, offspring generation relies on Simulated Binary Crossover and polynomial mutation [20]. Both operators are governed by a distribution index ($\eta$), independently set for each operator, which controls the spread of the generated offspring relative to their parents, governing the balance between exploration of the design space and exploitation of already-identified promising regions. Specifically, large values of $\eta$ concentrate the offspring distribution tightly around the parent solutions, favoring exploitation, while small values spread the offspring more broadly, favoring exploration. Values in the range $\eta \in [15, 20]$ are widely adopted as baseline settings for real-coded multi-objective evolutionary algorithms, as they provide an effective balance between these two competing objectives [20, 21], consistently with the default configurations of *pymoo* [3].

Evolutionary algorithms operating in non-convex design spaces provide no formal guarantee of convergence to the globally optimal Pareto front [19]; the identified front may constitute a near-optimal approximation of the true set of Pareto-optimal solutions. To characterize the degree of convergence achieved and assess the generalizability of the optimized designs beyond the specific conditions of the optimization, we performed two tests.

**Perturbation ensemble validation.** To assess the bias toward the fixed perturbation ensemble introduced during optimization, $\mathrm{FOM_R}$ of each Pareto-front individual was re-evaluated over an independent ensemble of $N = 500$ perturbed realizations, disjoint from the $N = 40$ set used during optimization. This sample size is sufficiently large to yield a statistically stable estimate of $\mathrm{FOM_R}$, in practice independent of the specific perturbation ensemble employed. A slight increase in $\mathrm{FOM_R}$ upon re-evaluation is statistically expected and serves as an indirect indicator of algorithmic convergence toward the fixed perturbation ensemble: a converged algorithm would have specifically exploited the fixed perturbation set, so that re-evaluated points exhibit higher $\mathrm{FOM_R}$ than their original counterparts, that is, they are dominated by the original Pareto-front individuals in the objective space. Re-evaluated points dominating the original front would instead signal lack of convergence. At the same time, the limited magnitude of this increase confirms that the optimized designs retain genuine structural robustness beyond the fixed perturbation ensemble, and that the overall sensitivity–robustness trade-off is preserved, as shown in Figs. S4b, S5b, S6b.

**Residual hypervolume** To quantitatively monitor algorithmic convergence across generations and enable direct comparability across independent optimization runs we tracked the residual hypervolume [22], defined as the fractional area of the bounded objective space not yet dominated by the current Pareto front: $\mathrm{FOM_R} \in [0\%, 100\%]$ and $S_\mathrm{I} \in [0,\ 14500]\ \mathrm{RIU}^{-1}$, where the upper bound on $S_\mathrm{I}$ was chosen to strictly exceed the maximum sensitivity observed across all evaluated individuals. This metric decreases monotonically as the Pareto front evolves toward a more favorable sensitivity–robustness trade-off, as discussed later in see Figs. S4a, S5a, S6a.

### Multi-stage optimization procedure

In the following, we detail the multi-stage optimization procedure that generated the final Pareto front presented in Fig. 2a of the main text. Each stage is initialized from the non-dominated solutions of the previous one, allowing the algorithm to build progressively upon already-acquired knowledge of the design space rather than restarting from uninformed initial conditions. In all stages, the population size was kept fixed at $N = 4000$ individuals, and the distribution index $\eta$ was set to the same value for both the Simulated Binary Crossover and the polynomial mutation operator.

**Stage 1: High-Exploration Initialization** ($\eta = 10$, 2000 generations)**.** NSGA-II was warm-started with $N = 4000$ individuals extracted from the single-objective Differential Evolution routine after 50

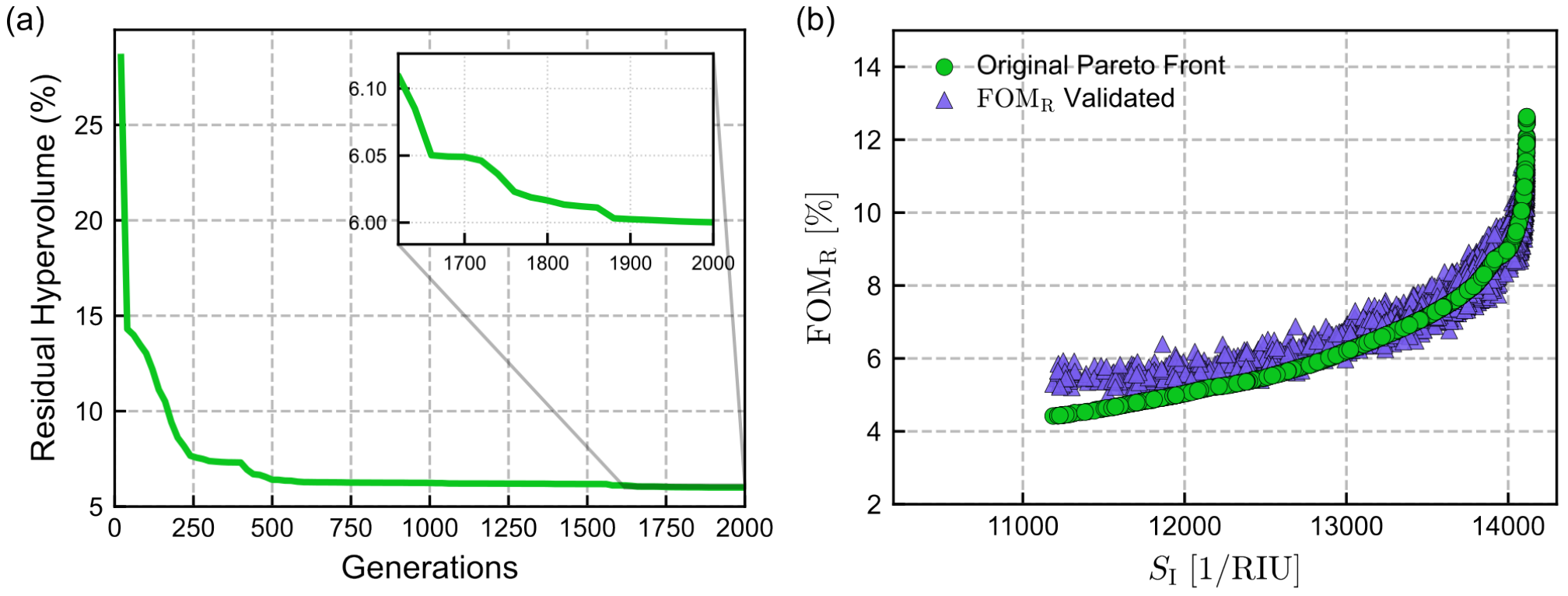


**Fig. S4.** Stage 1 optimization results ($\eta = 10$, 2000 generations). (a) Residual hypervolume as a function of generation number. The inset shows an enlarged view of the final 500 generations.(b) Perturbation ensemble validation of the Stage 1 Pareto front. For each structure of the original Pareto front (circles), FOM$_R$ has been re-evaluated over an independent ensemble of $N = 500$ perturbed realizations disjoint from the $N = 40$ set used during optimization (triangles).

generations, at a stage where the population already contained high-sensitivity candidates with a baseline optical resonance near the 60° target working angle, while retaining a high degree of parametric diversity. This initialization avoids the computational overhead of a purely random search phase while preserving broad exploration capability. The low value of $\eta = 10$ was selected to favor broad exploration of the design space, allowing offspring to spread widely across the parameter space. As shown in Fig. S4a, the residual hypervolume decreases sharply in the early generations before transitioning into a largely flat plateau, indicating that the algorithm reached a state of stagnation in which further generations at this $\eta$ value yield no significant improvement, with only a marginal residual convergence visible in the final 500 generations as shown in the inset of Fig. S4a. The perturbation ensemble validation (Fig. S4b) confirms that the sensitivity–robustness trade-off identified during optimization is preserved upon re-evaluation on an independent ensemble. However, the re-evaluated FOM$_R$ marginally outperforms the original optimization estimates for several Pareto-front individuals, which, consistently with the discussion above, indicates incomplete convergence at this stage. These observations motivate the transition to a higher $\eta$ regime in Stage 2, drawing inspiration from self-adaptive evolutionary strategies [21].

**Stage 2: Targeted Exploitation** ($\eta = 23$ and $\eta = 30$). The $N = 4000$ non-dominated individuals from Stage 1 were used as the initial population for two parallel optimization branches, each targeting a shift from broad exploration toward localized exploitation via higher distribution indices: $\eta = 23$ (500 generations) and $\eta = 30$ (300 generations). The fixed set of N × L percentage thickness variations was changed with respect to Stage 1 to disrupt potential stagnation and restore statistical dispersion within the objective space, restoring the algorithm's capability to escape local optima.

As shown in Fig. S5a, both branches successfully reduce the residual hypervolume beyond the Stage 1 plateau, with a rate of convergence markedly higher than the marginal improvement visible in the inset of Fig. S4a, and both resulting Pareto fronts dominate the Stage 1 front across the full sensitivity–robustness spectrum (not shown). The $\eta = 30$ branch achieves a lower final residual hypervolume, resulting in a Pareto front that outperforms $\eta = 23$ in the high-robustness regime; conversely, the $\eta = 23$ branch yields superior solutions at high sensitivity. The two Pareto fronts intersect, indicating that neither configuration strictly dominates the other across the full sensitivity–robustness spectrum. The perturbation ensemble validation test (Fig. S5b) shows that the re-evaluated FOM$_R$ lies above the original Pareto front for all individuals in both branches, consistently with the expected degradation upon re-evaluation at this level of convergence. Notably, the orange triangles associated with $\eta = 30$ exhibit a markedly larger deviation from the original front compared to those of $\eta = 23$, indicating a more pronounced specificity to the set of percentage thickness variations, as expected given the higher value of $\eta$. The intersection of the two fronts motivate the merging strategy adopted in Stage 3, where an

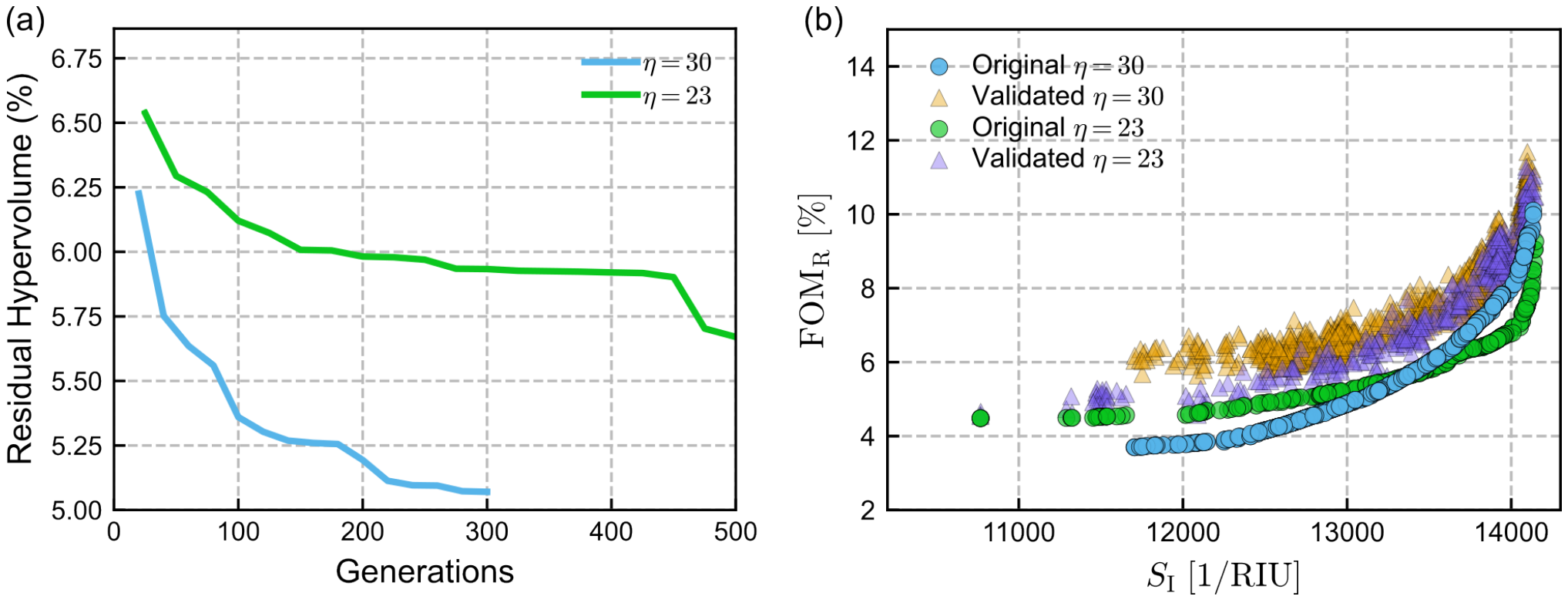


**Fig. S5.** Stage 2 optimization results. (a) Residual hypervolume as a function of generation number for the two parallel optimization branches, $\eta = 23$ (green, 500 generations) and $\eta = 30$ (light blue, 300 generations). (b) Perturbation ensemble validation of the Stage 2 Pareto fronts. Original Pareto front individuals are shown as circles ($\eta = 23$, green; $\eta = 30$, light blue) and their corresponding re-evaluated $\mathrm{FOM_R}$ as triangles ($\eta = 23$, purple; $\eta = 30$, orange).

intermediate distribution index is employed to consolidate the best solutions from both branches.

**Stage 3: Final Convergence and Consolidation** ($\eta = 28$, 1000 generations). The populations from the $\eta = 23$ and $\eta = 30$ branches were merged into a combined pool, from which an initial population of $N = 4000$ individuals was extracted via standard non-dominated sorting, progressively including subsequent Pareto ranks until the target size was reached. A new random perturbation ensemble and an intermediate distribution index ($\eta = 28$) were assigned.

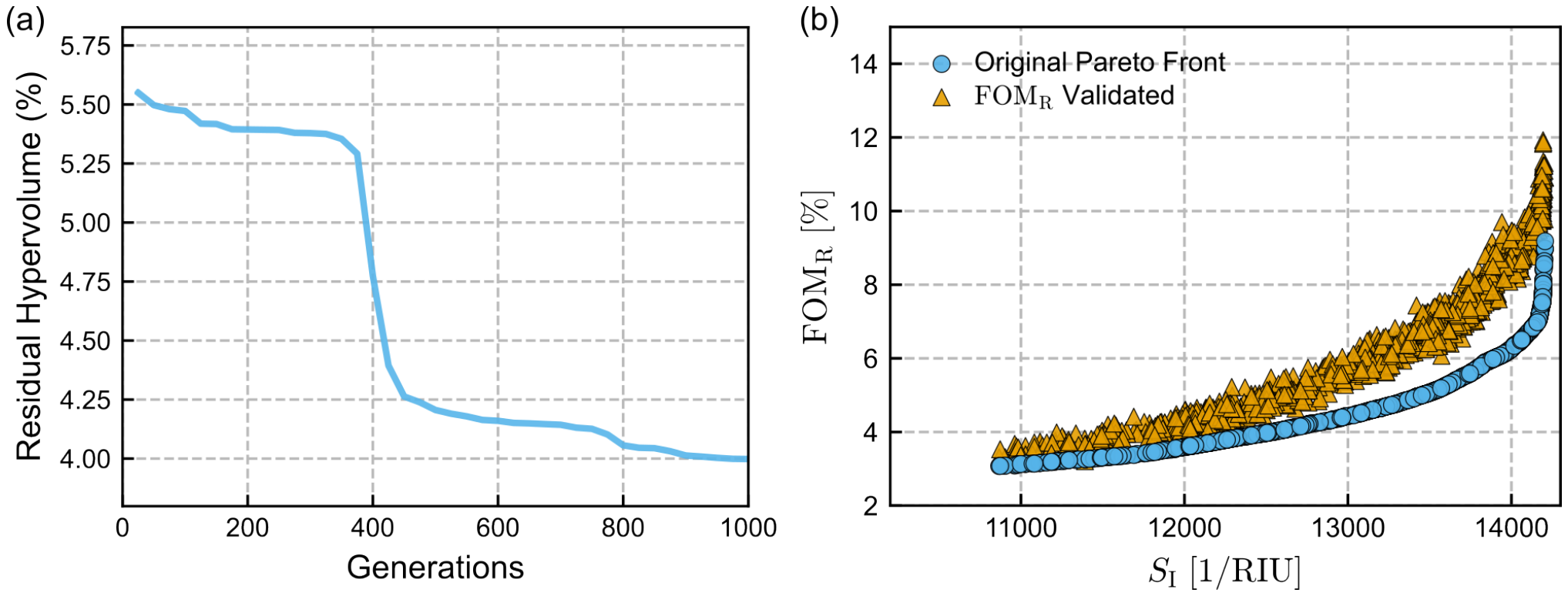


**Fig. S6.** Stage 3 final optimization results ($\eta = 28$, 1000 generations). (a) Residual hypervolume as a function of generation number. (b) perturbation ensemble validation of the final Pareto front. Original Pareto front structures (circles) and their corresponding re-evaluated $\mathrm{FOM_R}$ (triangles).

As shown in Fig. S6a, the residual hypervolume undergoes a further reduction of approximately 1.5 % over the 1000 generations. The perturbation ensemble validation test (Fig. S6b) shows that the re-evaluated $\mathrm{FOM_R}$ tracks the original Pareto front with only a minimal and uniform degradation, in contrast to the pronounced deviation observed for the $\eta = 30$ branch in Stage 2. These results confirm the genuine structural robustness of the final designs and validate a posteriori the choice of $N = 40$ perturbed realizations as a computationally efficient yet statistically sufficient ensemble size for the optimization. The resulting set of non-dominated solutions constitutes the Pareto front presented in Fig. 2a of the main manuscript.



\end{document}

%% file: Figure1.pdf_tex
\begingroup%
  \makeatletter%
  \providecommand\color[2][]{%
    \errmessage{(Inkscape) Color is used for the text in Inkscape, but the package 'color.sty' is not loaded}%
    \renewcommand\color[2][]{}%
  }%
  \providecommand\transparent[1]{%
    \errmessage{(Inkscape) Transparency is used (non-zero) for the text in Inkscape, but the package 'transparent.sty' is not loaded}%
    \renewcommand\transparent[1]{}%
  }%
  \providecommand\rotatebox[2]{#2}%
  \newcommand*\fsize{\dimexpr\f@size pt\relax}%
  \newcommand*\lineheight[1]{\fontsize{\fsize}{#1\fsize}\selectfont}%
  \ifx\svgwidth\undefined%
    \setlength{\unitlength}{265.60903931bp}%
    \ifx\svgscale\undefined%
      \relax%
    \else%
      \setlength{\unitlength}{\unitlength * \real{\svgscale}}%
    \fi%
  \else%
    \setlength{\unitlength}{\svgwidth}%
  \fi%
  \global\let\svgwidth\undefined%
  \global\let\svgscale\undefined%
  \makeatother%
  \begin{picture}(1,0.4910441)%
    \lineheight{1}%
    \setlength\tabcolsep{0pt}%
    \put(0,0){\includegraphics[width=\unitlength,page=1]{Figure1.pdf}}%
    \put(0.20234581,0.46126682){\color[rgb]{0,0,0}\makebox(0,0)[lt]{\lineheight{0}\smash{\begin{tabular}[t]{l}z\end{tabular}}}}%
    \put(0,0){\includegraphics[width=\unitlength,page=2]{Figure1.pdf}}%
    \put(0.21252886,0.08173441){\color[rgb]{0,0,0}\makebox(0,0)[lt]{\lineheight{0}\smash{\begin{tabular}[t]{l}$\theta$\end{tabular}}}}%
    \put(-0.00198541,0.20910268){\color[rgb]{0,0,0}\makebox(0,0)[lt]{\lineheight{0}\smash{\begin{tabular}[t]{l}n\textsubscript{low}\end{tabular}}}}%
    \put(-0.00198541,0.16646079){\color[rgb]{0,0,0}\makebox(0,0)[lt]{\lineheight{0}\smash{\begin{tabular}[t]{l}n\textsubscript{high}\end{tabular}}}}%
    \put(-0.00198541,0.07922344){\color[rgb]{0,0,0}\makebox(0,0)[lt]{\lineheight{0}\smash{\begin{tabular}[t]{l}n\textsubscript{sub}\end{tabular}}}}%
    \put(-0.00198541,0.35610478){\color[rgb]{0,0,0}\makebox(0,0)[lt]{\lineheight{0}\smash{\begin{tabular}[t]{l}n\textsubscript{sup}\end{tabular}}}}%
    \put(0,0){\includegraphics[width=\unitlength,page=3]{Figure1.pdf}}%
    \put(0.90239958,0.28989253){\color[rgb]{0.04705882,0.77647059,0.12156863}\makebox(0,0)[lt]{\lineheight{0}\smash{\begin{tabular}[t]{l}A\end{tabular}}}}%
    \put(0.849238,0.25740936){\color[rgb]{0.04705882,0.77647059,0.12156863}\makebox(0,0)[lt]{\lineheight{0}\smash{\begin{tabular}[t]{l}$\Gamma_\theta$\\\end{tabular}}}}%
    \put(0,0){\includegraphics[width=\unitlength,page=4]{Figure1.pdf}}%
  \end{picture}%
\endgroup%